\documentclass[]{article}
\newlength{\topmarge}
\newlength{\lrmarge}
\newcommand{\hepth}[1]{hep-th/\,#1}

\newcommand{\journal}[4]{{\em #1~}#2\,(19#3)\,#4}

\newcommand{\sjnp}{\journal {Sov. J. Nucl. Phys.}}

\newcommand{\prl}{\journal {Phys. Rev. Lett.}}

\newcommand{\np}{\journal {Nucl. Phys.}}
\newcommand{\pl}{\journal {Phys. Lett.}}

\makeatletter
\@addtoreset{equation}{section}
\makeatother
\renewcommand{\theequation}{\thesection.\arabic{equation}}
\newcommand{\es}{\\[3mm]}
\newcommand{\eq}{\begin{equation}}
\newcommand{\eqn}[1]{\label{#1}\end{equation}}
\newcommand{\ba}{\begin{array}}
\newcommand{\ea}{\end{array}}

\newcommand{\Lac}{\displaystyle{\biggl\{}}
\newcommand{\Rac}{\displaystyle{\biggr\}}}
\newcommand{\LP}{\displaystyle{\Biggl(}}
\newcommand{\RP}{\displaystyle{\Biggr)}}
\newcommand{\lp}{\left(}
\newcommand{\rp}{\right)}

\newcommand{\lac}{\left\{}
\newcommand{\rac}{\right\}}

\renewcommand{\a}{\alpha}
\renewcommand{\b}{\beta}
\renewcommand{\d}{\delta}
\newcommand{\e}{\varepsilon}
\newcommand{\eb}{{\bar\varepsilon}}
\newcommand{\f}{\phi}

\renewcommand{\k}{\kappa}
\renewcommand{\l}{\lambda}
\newcommand{\lb}{{\bar\lambda}}
\newcommand{\Lb}{{\bar\Lambda}}
\renewcommand{\L}{\Lambda}
\newcommand{\m}{\mu}
\newcommand{\n}{\nu}

\renewcommand{\r}{\rho}
\newcommand{\s}{\sigma}

\renewcommand{\t}{\theta}

\renewcommand{\AA}{{\mathcal A}}
\newcommand{\BB}{{\mathcal B}}

\newcommand{\EE}{{\mathcal E}}
\newcommand{\FF}{{\mathcal F}}

\newcommand{\II}{{\mathcal I}}

\newcommand{\LL}{{\mathcal L}}
\newcommand{\MM}{{\mathcal M}}

\newcommand{\RR}{{\mathcal R}}

\newcommand{\fb}{{\bar\f}}
\newcommand{\da}{{\dot{\a}}}
\newcommand{\db}{{\dot{\b}}}

\newcommand{\vf}{{\varphi}}

\newcommand{\complex}{{\kern .1em {\raise .47ex
                       \hbox {$\scriptscriptstyle |$}}
                       \kern -.4em {\rm C}}}
\newcommand{\real}{{{\rm I} \kern -.19em {\rm R}}}
\newcommand{\rational}{{\kern .1em {\raise .47ex
                        \hbox{$\scripscriptstyle |$}}
                        \kern -.35em {\rm Q}}}
\renewcommand{\natural}{{\vrule height 1.6ex width
                         .05em depth 0ex \kern -.35em {\rm N}}}
\newcommand{\znatural}{{ {\rm Z}\kern -.34em {\rm Z}}}

\newcommand{\dint}{\displaystyle{\int}}

\newcommand{\pa}{\partial}

\newcommand{\dpad}[2]{{\displaystyle{\frac{\pa #1}{\pa #2}}}}

\newcommand{\dfrac}[2]{{\displaystyle{\frac{#1}{#2}}}}

\newcommand{\na}{\nabla}

\newcommand{\bstar}{\begin{petit} \noindent {\Large $\star$} }
\newcommand{\estar}{\end{petit}}

\newcommand{\ie}{{{\em i.e.}\ }}

\newcommand{\sbar}{{\bar \s}}

\newcommand{\gP}{\Psi}

\newcommand{\ksl}{{\kern.12em {\raise.3ex\hbox{/} \kern-.59em k}}}
\newcommand{\psl}{{\kern.12em {\raise.3ex\hbox{/} \kern-.59em p}}}
\newcommand{\dsl}{{\kern.1em {\raise.3ex\hbox{/} \kern-.58em \partial}}}
\newcommand{\Asl}{{\kern.12em {\raise.3ex\hbox{/} \kern-.77em A}}}
\newcommand{\Dsl}{{\kern.15em {\raise.3ex\hbox{/} \kern-.7em D}}}

\newcommand{\kzero}{{\k\kern-.45em \raisebox{2mm}{\tiny o}\kern.30em}}
\newcommand{\lkzero}{{\k\kern-.35em \raisebox{1.6mm}{\tiny o}\kern.30em}}

\newcommand{\bQ}{{\bar Q}}

\newcommand{\refequ}[1]{(\ref{#1})}

\newcommand{\bX}{{\bar X}}

\newcommand{\bd}{{\bar \d}}

\newcommand{\bfm}{{\mathbf m}}
\newcommand{\bfn}{{\mathbf n}}
\newcommand{\bfx}{{\mathbf x}}
\newcommand{\bfy}{{\mathbf y}}

\newcommand{\leff}{{\LL^{\mathrm{eff}}}}
\newcommand{\im}{\mathrm{Im}}
\newcommand{\bFF}{{\bar\FF}}
\newcommand{\bJ}{{\bar J}}
\newcommand{\bPi}{{\bar\Pi}}
\newcommand{\IIm}{{(\II^{-1})}}
\newcommand{\phm}{\phantom{-}}
\newcommand{\tintbf}[1]{{\dint d^3 \! #1 \,}}
\newcommand{\sud}{{$SU(2)$}}
\newcommand{\uone}{{$U(1)$}}
\newcommand{\sudr}{{$SU(2)_{\scriptstyle{\mathrm R}}$}}
\newcommand{\hf}{{\hat\f}}
\begin{document}
\renewcommand{\theequation}{\arabic{equation}}
\renewcommand{\thesubsection}{\arabic{subsection}}
\thispagestyle{empty}
\begin{flushright}
hep-th~/~9905194\\
LPENSL~-~Th 11/99
\end{flushright}
\vspace{5mm}

\begin{center}
{\Large\bf Computation of the Central Charge for the Leading Order of the
$N=2$ Super-Yang-Mills Effective Action} \\
\vspace{10mm}
{\large Sylvain Wolf~\footnote{Supported by
the Swiss National Science Foundation.}}\\
\vspace{5mm}
{\sl Laboratoire de Physique, Ecole Normale Sup\'erieure de Lyon}\\
{\sl 46, All\'ee d'Italie, 69364 Lyon - Cedex 07, France} \\
{\small e-mail:} {\ttfamily swolf@ens-lyon.fr} \\
\vspace{5mm}

\begin{abstract}
\noindent
The central charge in the $N=2$  Super-Yang-Mills theory plays an
essential role in the work of Seiberg and Witten as it gives
the mass spectrum of the BPS states of the quantum theory.
Our aim in this note is to present a direct computation
of this central charge for the leading order  
(in a momentum expansion) of the
effective action. We will consider the $N=2$  Super-Yang-Mills theory with 
gauge group $SU(2)$. The leading order of the effective action
is given by the same holomorphic function $\FF$ appearing in the
low energy $U(1)$ effective action.
\end{abstract}
\end{center}

\vspace{4mm}

\vspace{12mm}
\subsection*{Introduction }

Tremendous progress has been made these last years in the
understanding of the strong coupling regime of the $N=2$ 
Super-Yang-Mills (SYM) theory,
pioneered by the work of Seiberg and Witten~\cite{seiberg-witten} for the
case of a \sud\ gauge group. They succeeded in determining a non-perturbative
solution for the leading order of the low energy 
effective action, which is given, as explained in~\cite{seiberg},  
by one single  holomorphic function $\FF(\gP)$, where $\gP$ is the
chiral $N=2$ multiplet that contains the gauge field for the unbroken
\uone\  gauge group. Later, it has been
shown that this solution is in fact unique~\cite{magro-sachs-lor}.

One essential tool used to obtain these results 
is the central charge $Z$ appearing in the $N=2$ 
Super-Poincar\'e algebra, because it gives a lower bound for the
mass $M$ of the states of the theory, known as the BPS bound~\cite{bps}:
\eq
M\geq |Z|\ .
\eqn{intro00}
States that saturate this bound are called BPS-states.
In fact, it turns out that all the states of the classical 
theory saturate this bound and that this remains true at the quantum
level~\cite{witten-olive}.

For the  classical $N=2$ SYM action with gauge group \sud , 
a direct computation  of $Z$ leads to~\cite{witten-olive}: 
\eq
Z=a\ \lp n_e+\tau n_m\rp\ \ ,
\eqn{intro0}
where $\ a\ $ is the Higgs field expectation value,  
$\tau=\frac{4\pi i}{g^2}+\frac{\t}{2\pi}$ is
the complex coupling constant, including the gauge coupling constant $g$
and the $\t$-angle, and $n_e$ and $n_m$ are
the electric and magnetic numbers of the state, respectively.

For the  leading order of the low energy effective action, 
Seiberg and Witten~\cite{seiberg-witten} argued that the central charge 
must be given by 
\eq
Z=a\ n_e\ +\ a_D\ n_m\ ,
\eqn{intro1}
where $a_D=\dpad{\FF}{a}(a)$.
A first check of this formula has been obtained 
in~\cite{jaegher-dewit-kleijn-vandoren}, where only the bosonic
contributions to $Z$ have been considered.
Recently, a direct and complete computation of $Z$ 
has been performed in~\cite{alfredo} and agrees with~\refequ{intro1}.

However it remains to show that the expression~\refequ{intro1} 
is true for the high energy \sud\ effective action. 
Indeed, even if the central charge is expected to be a boundary term
and thus to receive contributions only from the massless \uone\ fields, 
this has not been proved so far. 
A first step in that direction has been achieved 
in~\cite{chalmers-rocek-vonunge}, where~\refequ{intro1} has been 
obtained by applying the BPS trick to the bosonic 
part of the leading order of the \sud\ effective action.

In this note, we will also consider the leading order of the \sud\ effective
action, for which we will perform a direct
computation of $Z$ by evaluating the Dirac bracket of two
supersymmetry charges. The result shows that~\refequ{intro1} is the 
correct expression. 

The plan of this note is as follows. 
In the first section, we introduce the notations
and expand the effective action defined by $\FF$ in components. 
In section 2, the supersymmetry
current is constructed by performing a local supersymmetry transformation. In 
section 3 we implement the Dirac bracket, which is the natural canonical
structure in the presence of fermions. 
Finally the central charge is obtained in section 4 by computing the 
Dirac bracket of two supersymmetry charges. 

This note being only a sketchy presentation of the
tools used to define the supercurrent and the canonical structure
and of the computation itself, a 
more detailed publication is in preparation~\cite{lor-alfredo-sylvain}.

\subsection{Field Content and Effective Action}

The pure $N=2$ SYM action with gauge group \sud\  
contains a triplet of $N=2$ chiral multiplets
$\gP^a$ in the adjoint representation of \sud,  labeled by the index $a$. 
The leading order of the effective Lagrangian is given
in terms of $\FF(\gP^a)$, 
with the same holomorphic function $\FF$ as in the low energy
effective action~\cite{seiberg-witten}, by:
\eq
\leff=\dfrac{1}{4\pi}\; \dint d^4\t\,\im \lp \FF(\gP)\rp
\ .
\eqn{eq1}
The classical Lagrangian can be recovered by letting
$\FF(\gP)=\dfrac{1}{2}\tau\gP^a\gP^a$, where 
$\tau=\frac{4\pi i}{g^2}+\frac{\t}{2\pi}$.

We then expand this Lagrangian into components choosing 
the Wess-Zumino gauge and eliminating the auxiliary fields.
The remaining fields are the gauge field $A^a_\m$, 
the \sudr -doublet~\footnote{
We denote by \sudr\ the \sud\ that rotates the two supersymmetry charges into
each other. Our conventions about the \sudr\ indices are the following:
$\vf_\bfn=\e_{\bfn\bfm}\vf^\bfm$ with $\e_{\bfm\bfn}=-\e_{\bfn\bfm}\ ,\
\e_{\bf 21}=1\ $; $\vf^\bfm=\e^{\bfm\bfn}\vf_\bfn$ with
$\e^{\bfm\bfn}=-\e^{\bfn\bfm}\ ,\
\e^{\bf 12}=1\ $; Finally, $(\vf^\bfn)^*={\bar\vf}_\bfn$, and, as a
consequence, $({\bar\vf}^\bfn)^*=-\vf_\bfn$.} of Weyl spinors $\l^{a\bfn\a}$ ($\bfn$ is the \sudr\ index
whereas $\a$ is the spinor index) and the complex scalar field $\f^a$, all
in the adjoint representation of \sud. 
$A^a_\m$ and  $\f^a$ are \sudr -singlets.

This gives
\eq
\leff=\dfrac{1}{8\pi}\im\LP\ba[t]{l} 
\FF^{ab}(\f)
\lac
-\dfrac{1}{4}F^a_{\m\n}F^{b\m\n}
-\dfrac{i}{8}\e^{\m\n\r\l}F^a_{\m\n}F^b_{\r\l}
-i\l^{a\bfn}\s^\m\nabla_\m\lb^b_\bfn
+\dfrac{1}{2}\na_\m\f^a\na^\m\fb^b
\right.\es \left. 
-D^{a\bfm\bfn} D^b_{\bfm\bfn}
+\dfrac{i}{2}\e^{acd}(\l^{c\bfn}\l^b_\bfn)\fb^d
-\dfrac{i}{2}\e^{acd}(\lb^{c\bfn}\lb^b_\bfn)\f^d
+\dfrac{1}{8}(\e^{acd}\f^c\fb^d)(\e^{bef}\f^e\fb^f)\rac
\es
-\dfrac{1}{4}\FF^{abc}(\f)F^a_{\m\n}\lp\l^{b\bfn}\s^{\m\n}\l^c_\bfn\rp
-\dfrac{1}{6}\FF^{abcd}(\f)
\lp\l^{a\bfm}\l^{b\bfn}\rp
\lp\l^c_\bfm\l^d_\bfn\rp \RP\ ,
\ea
\eqn{eq4}
where $\e^{abc}$ are the structure constants of \sud\ and
$\FF^{a_1\cdots a_n}(\f)$
denotes the $n$-th derivative of $\FF$ with respect
to $\f^{a_1}\cdots \f^{a_n}$.
For further use, let us introduce
the notation  $\II^{a_1\cdots a_n}$ and $\RR^{a_1\cdots a_n}$
for the imaginary and real part of $\FF^{a_1\cdots a_n}$,
respectively.
Finally, we use the condensed notation
\eq
D^{a\bfn\bfm}=
-\frac{i}{4}\IIm^{ac}\lp
\FF^{bcd}(\l^{b\bfn}\l^{d\bfm})
+\bFF^{bcd}(\lb^{b\bfn}\lb^{d\bfm})\rp\ ,
\eqn{eq4b}
where $\IIm^{ab}$ denotes the inverse of $\II^{ab}$,
\ie $\IIm^{ac}\II^{cb}=\d^{ab}$. 

From now on, it is implicitly understood that 
$\FF$ and its derivatives (as well as $\II, \RR, \IIm$ and so on)
are to be evaluated at $\f$,
except when otherwise stated. 

\subsection{Supersymmetry  Transformations and Currents}

The Lagrangian~\refequ{eq4}
is invariant under the global on-shell $N=2$
supersymmetry transformations~\cite{n2susy}
(with $\d=\e^{\bfn\a}\d_{\bfn\a}-\eb^{\bfn\da}\bd_{\bfn\da}$):
\eq
\lac\ba{lcl}
\d A^a_\m &=&
\phm\e^\bfn\s^\m\lb^a_\bfn
+\l^{a\bfn}\s^\m\eb_\bfn
\ ,\es
\d\f^a &=&
\phm 2\e^\bfn\l^a_\bfn
\ ,\es
\d\l^{a\bfn} &=&
-\frac{1}{2}\e^\bfn\s^{\m\n}F^a_{\m\n}
+i\na_\m\f^a\eb^\bfn\sbar^\m
-2D^{a\bfn\bfm}\e_\bfm
-\frac{i}{2}\e^{abc}\f^b\fb^c\e^\bfn
\ .
\ea\right.
\eqn{eq5}
 To compute the supersymmetry  current, we perform a local
supersymmetry transformation, defined by
$\d_{\mathrm{local}}=-\eb^{\bfn\da}(x)\bd_{\bfn\da}$
(and similarly for $\d_{\bfn\a}$). 
Since the action is invariant under a global supersymmetry transformation
(which means that the Lagrangian 
transforms as a total derivative), we have, under a local transformation:
\eq
\d_{\mathrm{local}}\leff = -i\pa_\m\eb^{\bfn\da} \bJ^\m_{\bfn\da} 
+ \pa_\m V^\m\ ,
\eqn{eq6}
which {\em defines\ } the supersymmetry current $\bJ^\m_{\bfn\da}$.

{\bf Remark}:
In fact, this relation defines
the supersymmetry current $\bJ^\m_{\bfn\da}$ and the total divergence
$V^\m$ only up to  an {\em improvement term\,} 
$\bX^{[\m\n]}_{\bfn\da}$ since
${{\bJ}{}'}^\m_{\bfn\da}=\bJ^\m_{\bfn\da}+i\pa_\n \bX^{[\m\n]}_{\bfn\da}$ and 
${V'}^\m=V^\m-\eb^{\bfn\da}\pa_\n \bX^{[\m\n]}_{\bfn\da}$ satisfies
\refequ{eq6} as well.

This computation leads to: 
\eq
\bJ_{\bfn}^\m =\ba[t]{l}
\phm  \dfrac{i}{8\pi} \II^{ab}\lp
-\sbar_\n\l^a_\bfn \lp F^{b\m\n}-\frac{i}{2}\e^{\m\n\r\l}F^b_{\r\l}\rp
+\sbar^\n\s^\m\lb^a_\bfn\na_\n\f^b
-\frac{1}{2}\sbar^\m\l^a_\bfn \e^{bcd}\f^c\fb^d\rp
\es
+\dfrac{i}{16\pi}\bFF^{abc}\lp\lb^{a\bfm}\lp
\l^b_{\bfn}\s^\m\lb^c_{\bfm}\rp\rp
+\dfrac{i}{48\pi}\FF^{abc}\lp\sbar^\m\l^{a\bfm}\lp
\l^b_\bfn\l^c_\bfm\rp\rp
\es
+i\pa_\n \bX_{\bfn}^{[\m\n]}\ .
\ea
\eqn{eq8}
The improvement term $\bX_{\bfn}^{[\m\n]}$
will be fixed explicitly in the next section, by imposing that the canonical
supersymmetry charge generates the supersymmetry transformation
laws~\refequ{eq5} we started with.

\subsection{Canonical  Structure and Dirac Bracket}

To compute the supersymmetry algebra satisfied by the supersymmetry
charges, we first have to implement the canonical structure of the model.
Let us start by defining the canonical conjugate momenta of the dynamical 
fields (all fields excepted $A^a_0$): 
\eq\ba{l}
\ba[b]{lcccllcccl}
\Pi^a&=&\dpad{\leff}{\pa_0\f^a}&=&\phm\dfrac{1}{16\pi}\II^{ab}\na^0\fb^b\ ,&
\bPi^a&=&\dpad{\leff}{\pa_0\fb^a}&=&\phm\dfrac{1}{16\pi}\II^{ab}\na^0\f^b
\ ,\\[4mm]
\L^a_{\bfm\a}&=&\dpad{\leff}{\pa_0\l^{a\bfm\a}}&=&
-\dfrac{1}{16\pi}\bFF^{ab}\lp\s^0\lb^b_\bfm\rp_\a\ ,\hspace{2mm}&
\Lb^a_{\bfm\da}&=&\dpad{\leff}{\pa_0\lb^{a\bfm\da}}&=&
-\dfrac{1}{16\pi}\FF^{ab}\lp\l^b_\bfm\s^0\rp_\da\ ,
\ea
\\[4mm]
\ba{lcccl}
M^{aj}&=&\hspace{2.1mm}\dpad{\leff}{\pa_0 A^a_j}\hspace{2.1mm}&=&
\phm \dfrac{1}{8\pi}\lp
\II^{ab}E^{bj}
-\RR^{ab}B^{bj}
+\frac{i}{4}\FF^{abc}\lp\l^{b\bfn}\s^{0j}\l^c_\bfn\rp
+\frac{i}{4}\bFF^{abc}\lp\lb^{b\bfn}\sbar^{0j}\lb^c_\bfn\rp
\rp
\ ,\ea
\ea\eqn{eq10}
where $i,j,k,\ldots$ denote the space indices (and thus run from 1 to 3)
and we have set $E^{aj}= F^{aj0}$ and $B^{aj}= \frac{1}{2}\e^{0jkl}F^a_{kl}$
for the electric and magnetic non-abelian fields, respectively.
As usual, $A^a_0$ plays the role of a Lagrange multiplier and the associated
constraint is the Gauss law
\eq
0=\dpad{\leff}{A^a_0}=
\na_i M^{ai}+J^{a0}\ ,
\qquad\mbox{with}\qquad
J^{a0}=\im\lp\dfrac{i}{8\pi}\e^{abc}\FF^{bd}
\lp\l^{d\bfn}\s^0\lb^c_{\bfn}\rp
+2i\e^{abc}\f^b\Pi^c\rp\ .
\eqn{11c}
The canonical Poisson bracket~\footnote{We do not write explicitly 
the time argument of the fields, but it is always understood that both 
expressions appearing in a Poisson or a Dirac bracket
are evaluated at the same time.} is then defined by
\eq
\{\pi_m(\bfx),\vf^n(\bfy)\}_c\ =\ \d^n_m\d^{(3)}(\bfx-\bfy)\ ,
\eqn{eq11b}
where the $\pi_m$ denote the conjugate momenta to $\vf^m$.
In fact, the second line of~\refequ{eq10} defines two sets of constraints
$f^a_{\bfm\a}=0$ and ${\bar f}^a_{\bfm\da}=0$, where
$\ f^a_{\bfm\a}=\L^a_{\bfm\a}
+\frac{1}{16\pi}\bFF^{ab}\lp\s^0\lb^b_\bfm\rp_\a\ $ (and similarly for
${\bar f}^a_{\bfm\da}$). We thus apply the usual Dirac approach
for constrained systems~\cite{dirac}:
first, these constraints are easily recognized to be of the 
second class type; then, we construct the Dirac bracket, 
which ensures that the constraints
commute with any functions of $\vf^m$ and $\pi_m$.

More precisely, the procedure to construct the Dirac bracket 
is the following: let us denote by 
$f_i=0$ the constraints and by $m_{ij}$ the matrix
$\ \{f_i,f_j\}_c\ ,$ which is invertible 
for second class constraints.
The Dirac bracket of two functions $\AA$ and $\BB$ is then defined by
\[
\{\AA,\BB\}=\{\AA,\BB\}_c - \{\AA,f_i\}_c\ (m^{-1})^{ij}\ \{f_j,\BB\}_c\ ,
\]
and it can be indeed checked that it satisfies 
$\{\AA,f_i\}=0\ $ for all $\AA$ and $i$.

Applying this construction to the case under study means that 
we can indifferently use the spinor fields or their conjugate momenta and
leads to the following Dirac bracket: 
\eq
\ba{lcl}
\{M^{ai}(\bfx),A^b_j(\bfy)\}&=&\phm \d^{ab}\d^i_j\;\ \d^{(3)}(\bfx-\bfy)
\ ,\es
\{\Pi^a(\bfx),\f^b(\bfy)\}&=&\phm \d^{ab}\;\ \d^{(3)}(\bfx-\bfy)
\ ,\es
\{\Pi^a(\bfx),\l^{b\bfm\a}(\bfy)\}&=&\phm \dfrac{i}{2}
\lp\IIm^{bc}\FF^{acd}
\l^{d\bfm\a}\rp(\bfx)\;\ \d^{(3)}(\bfx-\bfy)
\ ,\es
\{\l^{a\a}_\bfn(\bfx),\lb^{b\bfm\da}(\bfy)\}&=&
\phm 8\pi i\IIm^{ab}(\bfx)\d^\bfm_\bfn\s^{0\a\da}
\;\ \d^{(3)}(\bfx-\bfy)
\ ,\es
\{\Pi^a(\bfx),\bPi^b(\bfy)\}&=&
-\dfrac{i}{32\pi}\lp\IIm^{cd}\FF^{ace}\bFF^{bdf}
\lp\l^{e\bfm}\s^0\lb^f_\bfm\rp\rp(\bfx)\;\ \d^{(3)}(\bfx-\bfy)
\ .
\ea
\eqn{eq12}
Note that the first two lines are the same as for the canonical bracket, but
we see in the next three lines how the Dirac construction modifies the relation
between the fields and the conjugate momenta for the fields
implied in the constraints.

\subsection{Supersymmetry and Central Charge}

To define a supersymmetry charge, we  have to choose one
current in the class defined by~\refequ{eq8}. As already
stated, the prescription is that we want that this charge
reproduces the transformation laws~\refequ{eq5} we started
with. Clearly, we define 
\eq
\bQ_{\bfm\da}=\tintbf{\bfx}\bJ^0_{\bfm\da}\ ,
\eqn{eq12b}
with $\bJ^0_{\bfm\da}$ given by~\refequ{eq8}
and impose that 
\eq
\{\bQ_{\bfm\da},\vf\}=i\bd_{\bfm\da}\vf\ ,
\eqn{eq12c}
paying attention to the boundary terms contributions. 
This leads to the particular
solution corresponding to $\bX_\bfn^{[\m\n]}=0$ in~\refequ{eq8},
and thus to the following supersymmetry charge: 
\eq
\bQ_\bfm
=i\ \tintbf{\bfx}\LP\ba[t]{l}
\sbar_j\l^a_\bfm \lp M^{aj}+\dfrac{1}{8\pi}\FF^{ab}B^{bj}\rp
+2\lb^a_\bfm\bPi^a
+\dfrac{i}{8\pi}\II^{ab}\sbar^{0j}\lb^a_\bfm\na_j\f^b
\es
-\dfrac{1}{16\pi}\II^{ab}\sbar^0\l^a_\bfm \e^{bcd}\f^c\fb^d
-\dfrac{1}{24\pi}\FF^{abc}\sbar^0\l^{a\bfn}\lp\l^b_\bfn\l^c_\bfm\rp\RP
\ .\ea
\eqn{eq13}

We can now obtain the central charge, defined by 
$\ Z=\frac{i}{8}\e^{\bfn\bfm}\e^{\da\db}\{\bQ_{\bfn\da},\bQ_{\bfm\db}\}\ $, 
by a tedious but straightforward computation which leads to 
\eq
Z=\tintbf{\bfx}\Lac
\pa_i\lp
\f^a M^{ai}
+\dfrac{1}{8\pi}\FF^a B^{ai}
-\dfrac{i}{32\pi}\FF^{ab}\lp\l^{a\bfn}\s^{0i}\l^b_\bfn\rp
\rp
-\f^a\lp\na_i M^{ai}+J^{a0}\rp
\Rac\ .
\eqn{eq21}
Now, taking into account the fact that the spinors
decrease faster than $r^{3/2}$ as $r\to\infty$ and that the last
term is the Gauss law, we obtain: 
\eq
Z=\tintbf{\bfx}\pa_i\LP\
\f^a M^{ai}
+\dfrac{1}{8\pi}\FF^a B^{ai}
\ \RP\ .
\eqn{eq22}
Let us then denote by $\ a\ $ the expectation value of the Higgs field 
and by $\hf^a$ the direction
(in the gauge space) where it points 
to~\footnote{$\hf^a$ is a real \sud\ triplet of unit norm, 
\ie $\hf^a\hf^a =1$.}. 
Thus, at spatial infinity, $\f^a(\bfx)=a\ \hf^a(\bfx)$. 
The direction defined by $\hf^a$ gives in fact the direction of the unbroken
\uone\ gauge symmetry that survives the Higgs phenomenon~\cite{witten-olive}. 
The \uone\ electric
and magnetic fields are thus defined by
$\EE^i=\hf^a E^{ai}$ and $\BB^i=\hf^a B^{ai}$ respectively, 
and we define in the same way $\MM^i=\hf^a M^{ai}$. 
Taking finally
into account the fact that the integrals of $\MM^i$ and of $\BB^i$
on the sphere at spatial infinity define (up to numerical factor)
the electric and magnetic numbers $n_e$ and $n_m$ of the state~\cite{witten},
we obtain the central charge in the form proposed by Seiberg and 
Witten~\cite{seiberg-witten} 
\eq
Z=a\ n_e\ +\ a_D\ n_m\ ,
\eqn{eq24}
where we have set~\footnote{In fact, we have
$a_D\equiv\dpad{\FF}{a}=\dpad{\f^a}{a}\dpad{\FF}{\f^a}=\hf^a\FF^a$ and
thus $\FF^a B^{ai}=a_D\hf^a B^{ai}=a_D\BB^i$.}  
$a_D=\dpad{\FF}{a}$.

\vspace{6mm}
{\bf Acknowledgments}: The author would like to thank M. Magro and I. Sachs
for introducing him to the work of Seiberg and Witten and for constant support 
during the computation, F. Delduc for extensive explanations about the 
Dirac bracket, and  L. O'Raifeartaigh and  A. Iorio 
for their kind hospitality
at the Dublin Institute for Advanced Studies
during the week where this paper has been achieved.

\end{document}